\input{aipcheck}

\documentclass[final,fix2col]{aipproc}

\layoutstyle{8x11double}

\begin{document}

\title{Identified particle transverse momentum spectra in $p$+$p$ and $d$+Au collisions at $\sqrt{s_{\mathrm {NN}}}$ = 200 GeV}



\author{Pawan Kumar Netrakanti (for STAR COLLABORATION)}{
  address={Variable Energy Cyclotron Centre, Kolkata}
}

\begin{abstract}
The transverse momentum ($p_{\mathrm T}$) spectra for identified 
charged pions, protons and anti-protons from $p$+$p$ and $d$+Au collisions 
are measured around  midrapidity ($\mid$y$\mid$ $<$ 0.5) over 
the range of 0.3 $<$ $p_{\mathrm T}$ $<$ 10 GeV/$c$ at 
$\sqrt{s_{\mathrm {NN}}}$ = 200 GeV. 
The charged pion and proton+anti-proton spectra at 
high $p_{\mathrm T}$ in $p$+$p$ collisions have been compared 
with the next-to-leading order perturbative 
quantum chromodynamic (NLO pQCD) calculations with a specific fragmentation 
scheme. The $p$/$\pi^{+}$ and $\bar{p}$/$\pi^{-}$
has been studied at high $p_{\mathrm T}$.
The nuclear modification factor ($R_{\mathrm dAu}$) shows that the 
identified particle Cronin effects around midrapidity are 
significantly non-zero for charged pions and to be even larger for protons at 
intermediate $p_{\mathrm T}$ (2 $<$ $p_{\mathrm T}$ $<$ 5 GeV/$c$).

\end{abstract}

\maketitle
\section{Introduction}
The study of identified hadron spectra at large transverse momentum
($p_{\mathrm T}$) in $p$+$p$ collisions can be used to test the predictions
from perturbative quantum chromodynamics (pQCD)~\cite{pqcd}. 
Comparisons between experimentally measured $p_{\mathrm T}$ spectra 
and theory can help to constrain the quark and gluon fragmentation functions.
Within the framework of pQCD, the expected initial-state nuclear effects
in $d$+Au collisions are multiple scattering (Cronin effect~\cite{cronin})
and shadowing of the parton distribution function. 
The study of the nuclear modification factor($R_{\mathrm dAu}$) will 
help us in understanding the nuclear effects involved in $d$+Au collisions.
The particle ratios at high $p_{\mathrm T}$
constrains particle production models and also gives unique
data on FF ratios, although extraction of this information is model-dependent.
For example, $p/\pi^{+}$ reflects the relative probability of a parton to
fragment into proton or pion at high $p_{\mathrm T}$~\cite{fnal_ratio}.
The above aspects have been discussed in this manuscript.

\begin{figure}
  \includegraphics[height=.32\textheight]{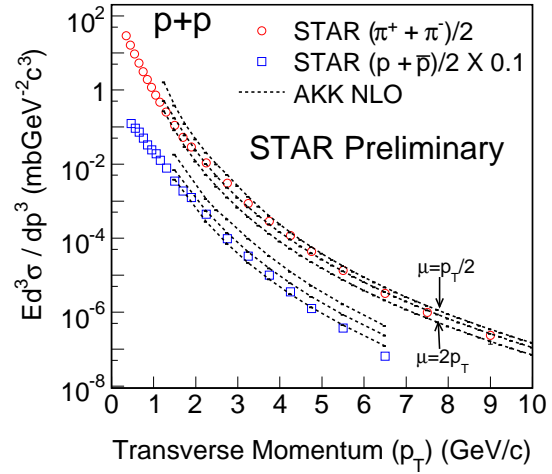}
  \caption{Midrapidity invariant yields for ($\pi^{+}$ +
$\pi^{-}$)/2 and ($p$+$\bar{p}$)/2 at high $p_{\mathrm T}$ for
minimum bias $p$+$p$ collisions compared to results from
NLO pQCD calculations using AKK~\cite{pqcd_akk} (PDF:
CTEQ6M) set of fragmentation functions. The calculations from AKK
are for three different factorization scales: $\mu$ = $p_{\mathrm T}/2$,
  $\mu$ = $p_{\mathrm T}$, and  $\mu$ = $2p_{\mathrm T}$.} 
\label{fig1}
\end{figure}

\section{Experiment and Analysis}
The detectors used in the present analysis are the Time Projection Chamber (TPC), 
the Time-Of-Flight (TOF) detector, a set of trigger detectors used for obtaining 
the minimum bias data, and the Forward Time Projection Chamber (FTPC) for the 
collision centrality determination in $d$+Au collisions in STAR experiment.
The details of the design and other characteristics of the detectors
can be found in Ref.~\cite{starnim}.
The data from TOF is used to obtain the identified hadron spectra 
for $p_{\mathrm T}<2.5$ GeV/$c$. The procedure for particle identification in TOF has
been described in Ref.~\cite{star_tof}.For $p_{\mathrm T}>2.5$ GeV/$c$, we use data
from the TPC. Particle identification at $p_{\mathrm T}$ in the TPC comes from 
the relativistic rise of the ionization energy loss (r$dE/dx$). 
Details of the method are described in Ref.~\cite{rdEdx}. 

\section{Comparison to NLO pQCD and model calculations}
In Fig.~\ref{fig1} we compare ($\pi^{+}$ + $\pi^{-}$)/2 and
($p$+$\bar{p}$)/2 yields in minimum bias $p$+$p$ collisions
at midrapidity for high $p_{\mathrm T}$ to those from NLO pQCD
calculations. The NLO pQCD results are based on calculations performed with 
{\it Albino-Kniehl-Kramer (AKK)} set of fragmentation functions~\cite{pqcd_akk}. 
We observe that our charged pion data for $p_{\mathrm T}$ $>$ 2 GeV/$c$ in $p$+$p$ 
collisions are reasonably well-explained by the NLO pQCD calculations using the AKK 
set of FFs. The calculations for the factorization scales of 
$\mu$ = $p_{\mathrm T}/2$, $\mu$ = $p_{\mathrm T}$, 
and $\mu$ = $2p_{\mathrm T}$ have been shown.
The combined proton and anti-proton yield in 
$p$+$p$ is lower compared to NLO pQCD calculations using AKK FFs for the 
factorization scale $\mu$ = $p_{\mathrm T}$. 
The ($p$+$\bar{p}$)/2 yield in $p$+$p$ collisions, however, is reasonably 
well-explained by AKK set of FFs for $\mu$ = 2$p_{\mathrm T}$.  
For the first time in  $p$+$p$ collisions we observe a reasonably good agreement 
between the NLO pQCD calculations (using AKK FFs) and data at high $p_{\mathrm T}$.
This reflects the importance of the flavor-separated measurements in
$e^{+}$+$e^{-}$ collisions in determining the FFs to baryons as used in AKK 
FFs calculation. 

\section{Nuclear modification factor}
\begin{figure}
  \includegraphics[height=.29\textheight]{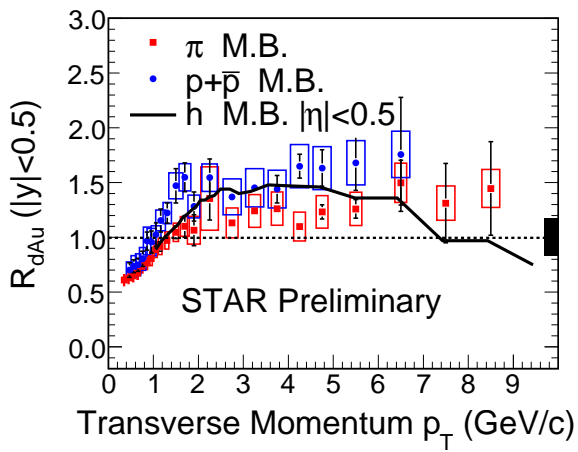}
  \caption{
Nuclear modification factor $R_{\mathrm dAu}$ for charged pions
($\pi^{+}+\pi^{-}$)/2 and $p$+$\bar{p}$ at $\mid$y$\mid$ $<$ 0.5
in minimum bias $d$+Au collisions. For comparison results on
inclusive charged hadrons (STAR) from Ref.~\cite{star_rdau}
at $\mid\eta\mid$ $<$ 0.5 are shown. The shaded band is the normalization 
uncertainty from trigger and luminosity in $p$+$p$ and $d$+Au collisions.} 
\label{fig2}
\end{figure}
The nuclear modification factor ($R_{\mathrm {dAu}}$) can be used to
study the effects of cold nuclear matter on particle production.
It is defined as a ratio of the invariant yields of the produced
particles in $d$+Au collisions to those in $p$+$p$ collisions scaled
by the underlying number of nucleon-nucleon binary collisions.
\begin{displaymath}
\nonumber
R_{\rm{dAu}}(p_{\rm T})\,=\,\frac{d^2N_{\rm{dAu}}/dy dp_{\rm T}}
{\langle N_{\rm {bin}}\rangle /\sigma_{\rm{pp}}^{\rm {inel}}\,d^2\sigma_{\rm{pp}}/dy dp_{\rm T}},
\end{displaymath}
where $\langle N_{\mathrm {bin}}\rangle$ is the average number
of binary nucleon-nucleon (NN) collisions per event, and
$\langle N_{\mathrm {bin}}\rangle /\sigma_{\rm{pp}}^{\mathrm {inel}}$
is the nuclear overlap function $T_A(b)$~\cite{star_rdau,star_pid}.
The $\sigma_{\rm{pp}}^{\rm {inel}}$ is taken to be 42 mb.

\begin{figure}
  \includegraphics[height=.36\textheight]{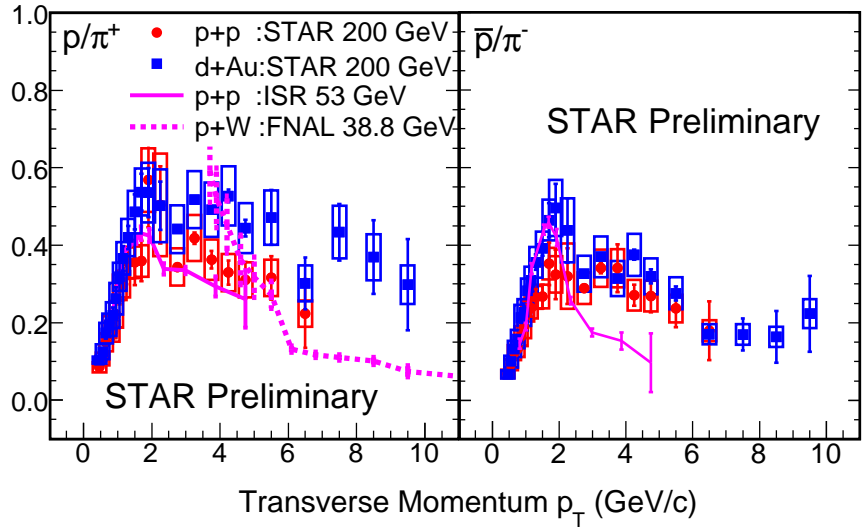}
  \caption{
Ratio of $p/\pi^{+}$ and $\bar{p}/\pi^{-}$ at midrapidity ($\mid$y$\mid$ $<$ 0.5) 
as a function of $p_{\mathrm T}$ in $p$+$p$ and $d$+Au minimum bias collisions.
For comparison the results from lower energies at ISR and FNAL
are also shown for $p/\pi^{+}$ and $\bar{p}/\pi^{-}$.
The errors represented by boxes are the point--to--point systematic.} 
\label{fig3}
\end{figure}
In Fig.~\ref{fig2} shows the $R_{\mathrm {dAu}}$ for charged pions 
(($\pi^{+}$+$\pi^{-}$)/2) and combined proton and anti-proton ($p$+$\bar{p}$) in 
minimum-bias collisions at $\mid$y$\mid$ $<$ 0.5. 
The $R_{\mathrm {dAu}}$ $>$ 1 indicates a slight enhancement of high $p_{\mathrm T}$ 
charged pions yields in $d$+Au collisions compared to binary collision scaled 
charged pion yields in $p$+$p$ collisions within the measured (y, $p_{\mathrm T}$)
range. The $R_{\mathrm {dAu}}$ for $p+ \bar{p}$ is again greater than unity for 
$p_{\mathrm T}$ $>$ 1.0 GeV/$c$ and is larger than that of the charged pions. 
The $R_{\mathrm {dAu}}$ results for identified particles has also been compared
to the inclusive charged hadrons. 
The uncertainty in determining the number of binary collisions in 
$d$+Au minimum-bias collisions is $\sim$5.3\%. 

\section{Particle ratio}
The $p/\pi^{+}$ and $\bar{p}/\pi^{-}$ at midrapidity as a function of $p_{\mathrm T}$
for $p$+$p$ and $d$+Au minimum bias collisions are shown in Fig.~\ref{fig3}.
At RHIC, the $p/\pi^{+}$ and $\bar{p}/\pi^{-}$
ratios increase with $p_{\mathrm T}$ up to 2 GeV/$c$ and then start
to decrease for higher $p_{\mathrm T}$ in both $p$+$p$ and $d$+Au
collisions.  The $\bar{p}/\pi^{-}$ ratio rapidly approaches a value of
0.2, which is also observed in $e^{+}$+$e^{-}$ collisions for both
quark and gluon jets~\cite{opal}. The $p/\pi^{+}$ ratios in $p$+$p$ collisions 
compare well with results from lower energy ISR and FNAL fixed target
experiments~\cite{isr53,fnal}, while $\bar{p}/\pi^{-}$ ratios at high
$p_{\mathrm T}$ have a strong energy dependence with larger values at
higher beam energies. 

\section{Summary}
We have measured the transverse momentum spectra for identified
charged pions, protons and anti-protons from $p$+$p$ and $d$+Au collisions
at $\sqrt{s_{\mathrm {NN}}}$ = 200 GeV around midrapidity ($\mid$y$\mid$ $<$ 0.5)
over the range of 0.3 $<$ $p_{\mathrm T}$ $<$ 10 GeV/$c$. For particle
identification we use the ionization energy loss and its relativistic
rise in the Time Projection Chamber and the Time-of-Flight in STAR.
The charged pions, combined proton and anti-proton spectra in $p$+$p$ and
collisions have been compared to calculations with the next-to-leading order
perturbative QCD calculations with a specific fragmentation scheme.
The NLO pQCD calculation explains the high $p_{\mathrm T}$ data for
charged pions reasonably well for $p_{\mathrm T}$ $>$ 2 GeV/$c$ in
$p$+$p$ collisions.  The $p$+$\bar{p}$ spectra are reasonably well-explained 
for the first time by NLO pQCD calculation using the AKK set of FFs with the 
factorization scale of $\mu$ = 2$p_{\mathrm T}$.  
An improved description of experimental data in RHIC's $p$+$p$ collisions by 
AKK FFs, which comes from NLO pQCD fits to the flavor separated $e^{+}$+$e^{-}$ 
data, is extremely interesting. These findings may provide a better foundation for
applications of jet quenching and quark recombination models to
explain the phenomena in A+A collisions in this $p_{\mathrm T}$ range.
Cronin effect around midrapidity is obsereved to be significantly
non-zero for pions, while the effect on proton and anti-proton spectra
is even larger at the intermediate $p_{\mathrm T}$ (2 $<$ $p_{\mathrm T}$ $<$ 5
GeV/$c$).  
The $p$/$\pi^{+}$ and $\bar{p}$/$\pi^{-}$ ratios have been studied at high 
$p_{\mathrm T}$ for $p$+$p$ and $d$+Au collisions. 
$p$/$\pi$ ratios peak at $p_{\mathrm T}\simeq2$ GeV/$c$ with a value of $\sim0.5$, 
and then decrease to $\sim0.2$ at high $p_{\mathrm T}$ with the possible
exception of the $p/\pi^{+}$ ratio in $d$+Au collisions. 

\begin{theacknowledgments}
We would like to thank Simon Albino for providing us the NLO pQCD results.
We also thank Werner Vogelsang and Stefan Kretzer for useful discussions.
\end{theacknowledgments}

\bibliographystyle{aipproc}   

\end{document}